\documentclass[sn-mathphys,Numbered]{sn-jnl}% Math and Physical Sciences Reference Style
\UseRawInputEncoding
\usepackage{anyfontsize}
\usepackage{multirow}%
\usepackage{amsmath,amssymb,amsfonts}%
\usepackage{amsthm}%
\usepackage{mathrsfs}%
\usepackage[title]{appendix}%
\usepackage{xcolor}%
\usepackage{textcomp}%
\usepackage{manyfoot}%
\usepackage{booktabs}%
\usepackage{algorithm}%
\usepackage{algorithmicx}%
\usepackage{algpseudocode}%
\usepackage{listings}%
\usepackage{graphicx}%

\usepackage{soulpos}
\ulposdef{\hlst}{%
    \rlap{\textcolor{yellow}{\rule[-.75ex]{\ulwidth}{2.5ex}}}%
    \rule[.45ex]{\ulwidth}{.1ex}%
}

\hypersetup{pdfstartview={FitH},pdfpagemode={UseNone},
            colorlinks,linkcolor=blue, citecolor=blue, urlcolor=blue,
            bookmarksopen=true, pdfnewwindow=true}
\usepackage[all]{hypcap}

\begin{document}
\title[Article Title]{Quantum Imaging Using Spatially Entangled Photon Pairs from a Nonlinear Metasurface}

\author[1]{\fnm{Jinyong} \sur{Ma}}
\equalcont{These authors contributed equally to this work.}

\author[1]{\fnm{Jinliang} \sur{Ren}}
\equalcont{These authors contributed equally to this work.}

\author[1,2]{\fnm{Jihua} \sur{Zhang}}

\author[3]{\fnm{Jiajun} \sur{Meng}}

\author[1]{\fnm{Caitlin} \sur{McManus-Barrett}}

\author[3,4]{\fnm{Kenneth B.} \sur{Crozier}}

\author*[1]{\fnm{Andrey A.} \sur{Sukhorukov}}\email{andrey.sukhorukov@anu.edu.au}

\affil[1]{\orgdiv{ARC Centre of Excellence for Transformative Meta-Optical Systems, Department of Electronic Materials Engineering, Research School of Physics}, \orgname{The~Australian National University}, \orgaddress{\street{}\city{Canberra}, \postcode{2600}, \state{ACT}, \country{Australia}}}

\affil[2]{\orgdiv{Songshan Lake Materials Laboratory}, \orgaddress{\street{}\city{Dongguan}, \postcode{523808}, \state{P. R. China}, \country{Australia}}}

\affil[3]{\orgdiv{ARC Centre of Excellence for Transformative Meta-Optical Systems, School of Physics}, \orgname{University of Melbourne}, \orgaddress{\street{}\city{Melbourne}, \postcode{3010}, \state{Victoria}, \country{Australia}}}

\affil[4]{\orgdiv{ARC Centre of Excellence for Transformative Meta-Optical Systems, Department of Electrical and Electronic Engineering}, \orgname{University of Melbourne}, \orgaddress{\street{}\city{Melbourne}, \postcode{3010}, \state{Victoria}, \country{Australia}}}

% \sloppy

\abstract{Nonlinear metasurfaces with subwavelength thickness 
were recently established as versatile platforms for the enhanced and tailorable generation of entangled photon pairs.
%have recently 
%revealed unlimited potential in facilitating the enhanced and tailorable generation of entangled photon pairs. 
The small dimensions and inherent stability of integrated metasurface sources are attractive for free-space applications in quantum communications, sensing, and imaging, yet this remarkable potential remained unexplored. 
%Yet the applications of photon pairs from metasurfaces remained unexplored, such as quantum communication, sensing and imaging. 
Here, we formulate and experimentally demonstrate the unique benefits and practical potential of nonlinear metasurfaces for quantum imaging at infrared wavelengths, facilitating an efficient protocol combining ghost and all-optical scanning imaging. The metasurface incorporates a subwavelength-scale silica metagrating on a lithium niobate thin film. %A distinguishing feature of the metasurface 
Its distinguishing feature is the capability to all-optically scan the photon emission angle in the direction across the grating simply by tuning the pump beam wavelength. Simultaneously, the photon emission is broad and anti-correlated along the grating direction, allowing for ghost imaging. Thereby, we reconstruct the images of 2D objects using just a 1D detector array in the idler path and a bucket detector in the signal path, by recording the dependencies of photon coincidences on the pump wavelength. Our results reveal new possibilities for quantum imaging with ultra-large field of view and improved imaging resolution as compared to photon pairs from conventional bulky crystals. The demonstrated concept can be extended to multi-wavelength operation and other applications such as quantum object tracking, paving the way for advancements in quantum technologies using ultra-compact nanostructured metasurfaces.}


\maketitle

\section{Introduction}
Nonlinear flat optics devices~\cite{maEngineering2024, kanAdvances2023, fedotovaSecondHarmonic2020d, Fedotova:2022-3745:ACSP, guoUltrathin2023, weissflogTunable2023, Solntsev:2021-327:NPHOT}, composed of nanostructured nonlinear materials with a sub-wavelength thickness,
%of only a few hundred nanometers, recently hold significant promise for 
were recently showcased as highly versatile, flexible and miniaturized photon pair sources based on spontaneous parametric down-conversion (SPDC). Notably, metasurfaces~\cite{HuangPlanar2020, neshevEnabling2023a, angelisNonlinear2020, nikolaevaDirectional2021} featuring nanostructures that support optical resonances can enhance and tailor the SPDC process beyond what is possible with conventional bulky nonlinear crystals. In recent years, it was experimentally demonstrated that nonlinear metasurfaces can facilitate the strongly enhanced generation of photon pairs~\cite{marinoSpontaneous2019a, parryEnhanced2021i, mazzantiEnhanced2022a, Santiago-Cruz:2021-4423:NANL}. Furthermore, the quantum photon state can be engineered to feature spectral~\cite{Santiago-Cruz:2022-991:SCI}, polarization~\cite{maPolarization2023a, weissflogNonlinear2024, jiaEntangled}, and spatial~\cite{Zhang:2022-eabq4240:SCA,zhangPhoton2023a} entanglement. These features appear highly attractive for free-space applications including quantum imaging, yet the fundamental potential of photon-pair sources with flat optics for any practical applications remained unexplored in previous studies.
%jinEfficient2021, 
%was not previously investigated.
%, such as quantum imaging. 

Quantum imaging systems based on spatially entangled photon pairs~\cite{moreauImaging2019, pittmanOptical1995, Padgett:2017-20160233:PTRSA, gilabertebassetPerspectives2019,shihQuantum2007c,cameronTutorial2024} potentially offer fundamental advantages over classical imaging techniques, allowing ultra-low photon flux operation~\cite{gilabertebassetPerspectives2019}, sensitivity beating shot-noise limit~\cite{bridaExperimental2010a,samantarayRealization2017}, higher security~\cite{malikQuantumsecured2012, magana-loaizaQuantum2019}, and improved resolution exceeding classical diffraction~\cite{botoQuantum2000, toninelliResolutionenhanced2019}. While it is still challenging to effectively replace classical approaches, new quantum imaging modalities show significant implications for real-world applications, such as quantum ghost imaging~\cite{morrisImaging2015, Padgett:2017-20160233:PTRSA, vegaPinhole2020, moreauResolution2018a, vegaMetasurfaceAssisted2021}, imaging with undetected photons~\cite{lemosQuantum2014, vegaFundamental2022, fuenzalidaResolution2022}, Hong-Ou-Mandel microscopy~\cite{ndaganoQuantum2022a}, adaptive optical imaging~\cite{cameronAdaptive2024}, 
and other schemes~\cite{zhouMetasurface2020b, lopaevaExperimental2013, Mitchell:2004-161:NAT, kalashnikovInfrared2016c,solntsevLiNbO32018}. Entangled photon pairs for quantum imaging can be generated from nonlinear crystals through SPDC. Traditionally the crystal thickness is at millimeter-scale such that the SPDC processes are subject to the longitudinal phase matching condition~\cite{santiago-cruzEntangled2021e, OkothMicroscale2019, guoUltrathin2023}.
The photon emission thus encompasses a narrow range of transverse momenta~\cite{gilabertebassetPerspectives2019} and significantly constrains the field of view for imaging~\cite{moreauResolution2018a}.
%under no usage of scanning techniques. 
More importantly, the photon pair sources from bulky crystals have limited tunability in all degrees of freedom of photons, thereby restricting the extensive potential and functionality of quantum imaging.

Here we reveal, for the first time to our knowledge, the unique benefits and practical potential of quantum imaging based on a nonlinear metasurface, facilitating an efficiently combined ghost and all-optical scanning imaging protocol at telecom wavelengths. We utilize the spatially entangled photon pairs generated from a metasurface where a subwavelength-scale silica grating is on top of a lithium niobate thin film of 300~nm thickness~\cite{Zhang:2022-eabq4240:SCA, weissflogDirectionally2024}. The nonlocal resonances supported by the metasurface result in two unique properties for the photon pair emission: (i) the angular emission is narrow in $y$- (orthogonal to the grating) yet broad in $z$-direction (parallel to the grating)~\cite{Zhang:2022-eabq4240:SCA} and (ii) the emission angle in $y$-direction can be all-optically scanned over a wide range by simply tuning the pump beam wavelength~\cite{weissflogDirectionally2024}. Using these properties, we realize a 2D imaging combining quantum ghost imaging in $z$-direction and optical scanning imaging in $y$-direction. Because the metasurface is sub-wavelength thick and contemporary fabrication capabilities allow for centimeter-scale apertures~\cite{huangSubwavelength2022,andrenLargeScale2020,sheLarge2018a}, our protocol presents opportunities to substantially expand the field of view and improve imaging resolution to the diffraction limit~\cite{vegaFundamental2022}. Furthermore, the optical scanning characteristic enables 2D infrared imaging using just a one-dimensional detector array and this principle can be extended to multi-color imaging. Pumping this passive metasurface with a frequency-comb source~\cite{shaltoutSpatiotemporal2019b} may also allow ultra-fast beam steering of biphoton emission, which is previously unattainable with bulky crystals and may open the door to more applications such as quantum object tracking and quantum LiDAR~\cite{lopaevaExperimental2013a, liuEnhancing2019, Mrozowski:2024-2916:OE, zhaoLight2022}.

% \comment{*raster-scanned~\cite{Padgett:2017-20160233:PTRSA}}
% Lastly, miniaturized light sources can further facilitate integrated quantum imaging systems.

%Because the metasurface is sub-wavelength thick and the present technology enables the fabrication of a metasurface with an aperture larger than a centimeter scale, our protocol opens the door to enlarging the field of view and improving the imaging resolution down to the diffraction limit. Furthermore, the optical scanning feature allows infrared imaging with only a 1D detector array and provides a practical potential for object tracking, which is impossible with bulky crystals. Finally, quantum imaging with the miniaturized light source can further facilitate the integrated quantum imaging systems. 

\begin{figure}[ht]
\centering
\includegraphics[width=0.78\linewidth]{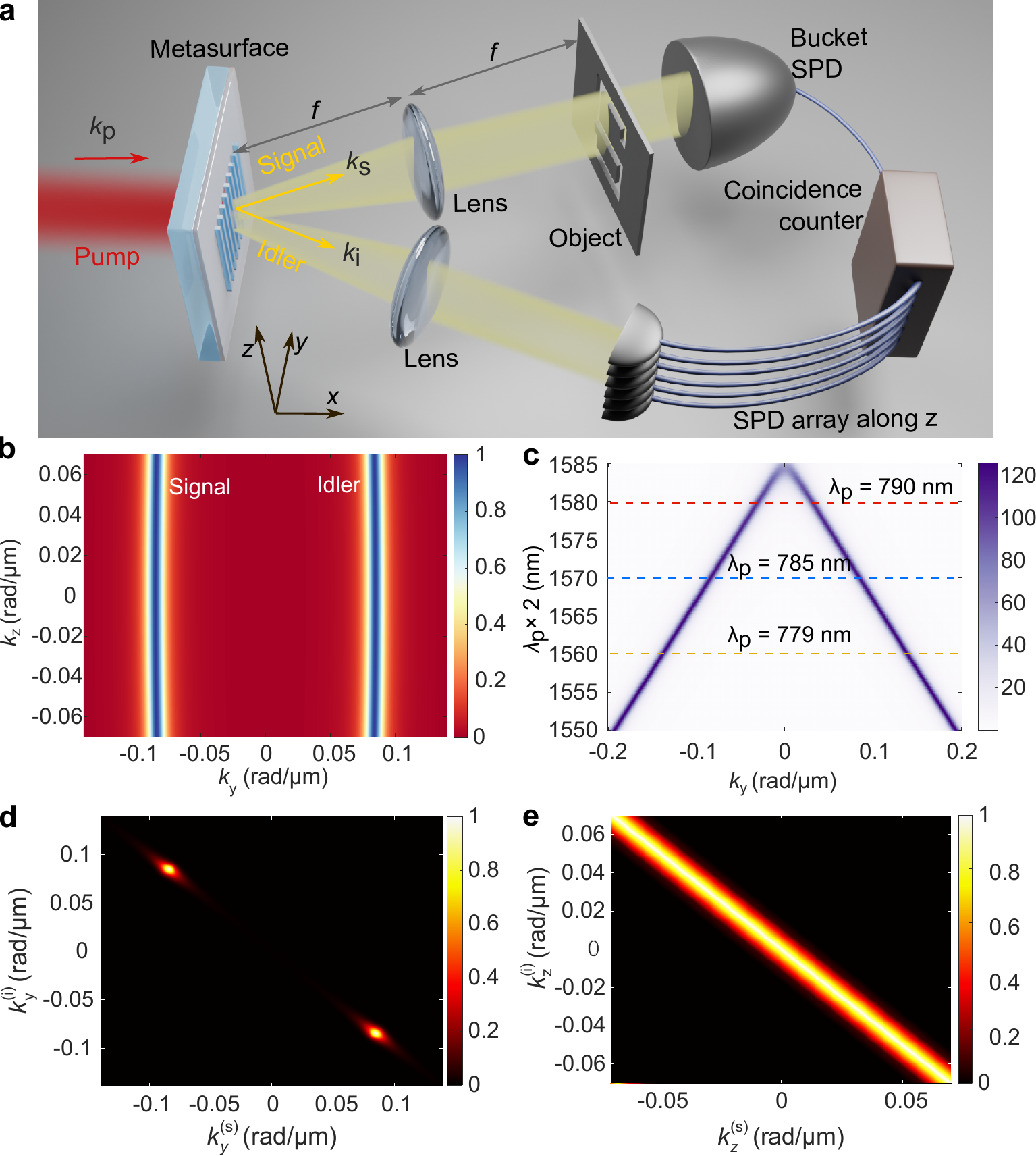}                                                   
\caption{\textbf{Concept.} \textbf{a,}~Quantum imaging protocol with photon pairs from a nonlinear metasurface. The metasurface, incorporating a silica grating atop a lithium niobate thin film, produces spatially entangled signal and idler photons where their emission direction along $y$-direction is tunable via pump laser wavelength. Signal photons passing through the object are collected using a bucket single-photon detector (SPD), while idler photons without interacting with the object are captured by a $z$-oriented 1D SPD array. We perform quantum ghost imaging and all-optical scanning in the $z$- and $y$-direction respectively, achieving 2D imaging of the object.
 \textbf{b,}~Simulated photon emission rate vs. transverse wavevector components $k_y$ and $k_z$. The photon emission is continuous and broad along $k_z$ while narrow in the $k_y$-direction. The pump wavelength is $1570/2$~nm.
 \textbf{c,}~Predicted photon-pair rate vs. $\lambda_{\rm p}\times 2$ and $k_y$ at $k_z=0$. The emission angle of photon pairs is optically controllable via the pump wavelength $\lambda_{\rm p}$, allowing optical scanning imaging. 
  \textbf{d-e,}~Biphoton joint spatial intensity along $k_y$ and $k_z$. The spatial anti-correlation of photon pairs in $y$-direction enables the deterministic separation of the signal and idler photons. In the $z$-direction, the signal and idler photons are spatially anti-correlated with continuous emission, allowing quantum ghost imaging. Here $\lambda_{\rm p} = 1570/2$ nm is used.
}
\label{fig:1}
\end{figure}

\section{Results}
\subsection{Concept}
The proposed quantum imaging protocol is schematically represented in Fig.~\ref{fig:1}\textbf{a}. A pair of spatially correlated photons (called signal and idler) is generated from a nonlinear metasurface~\cite{Zhang:2022-eabq4240:SCA} and is spatially split into two optical paths. According to the general principles of ghost imaging~\cite{Padgett:2017-20160233:PTRSA}, the target object is placed in an optical path where the arrival of the signal photon is measured with a bucket single-photon detector that does not by itself provide information on the spatial features of the object. The spatially-resolved detection of the idler photon in the other optical path, which never passes through the object, can allow the reconstruction of the object image using spatial correlations of the photon pair. Traditionally a 2D camera incorporating single-photon detector arrays or electronically reconfigured 2D light modulators are needed in the idler path to reconstruct 2D objects. 
%However single-photon 2D camera has been technically challenging in the telecom wavelength domain. 
Here the 2D camera is replaced with a simple 1D detector array in the idler arm to accomplish the 2D imaging. We anticipate that an imaging protocol based on wavelength-dependent scattering~\cite{shinSinglepixel2016} may be potentially implemented with nonlinear metasurfaces to further reduce the number of detector pixels.

% The nonlinear metasurface incorporates a sub-wavelength-thick lithium niobate film covered by a silica linear grating.

The metasurface, comprising a lithium niobate thin film covered by a silica meta-grating, supports dual nonlocal optical resonances with distinct angular dispersions in two orthogonal directions~\cite{Zhang:2022-eabq4240:SCA}: nearly flat along the grating orientation ($z$-direction) while quadratic in the y-direction (i.e. orthogonal to grating). Consequently, photon pairs emitted from the metasurface exhibit a narrow emission angle in $y$- yet broad in $z$-direction, as depicted in Fig.~\ref{fig:1}\textbf{b}. The strongly angular-dependent resonances along $y$-direction enable the tuning of the photon emission angle, determined by the transverse wavevector component
$k_y$, by simply adjusting the optical pump wavelength, as shown in Fig.~\ref{fig:1}\textbf{c}. This tuning feature has been experimentally demonstrated in co- and counter-propagating photon-pair emission from similar metasurfaces~\cite{weissflogDirectionally2024}. The tuning range can potentially cover the whole $k$-space, allowing all-optical scanning imaging with an ultra-large field of view. Because the entangled pairs of signal and idler photons acquire transverse momenta that are nearly equal in magnitude and have opposite signs for a normally incident plane-wave pump light, i.e.,  $k_{y,z}^{(s)} \simeq - k_{y,z}^{(i)}$,  they are anti-correlated at a narrow $k_y$ range (Fig.~\ref{fig:1}\textbf{d}), while continuously anti-correlated along $z$-direction (Fig.~\ref{fig:1}\textbf{e}). This characteristic allows for deterministic separation of signal and idler photons in the $y$-direction and quantum ghost imaging of a narrow object slice oriented along the $z$-direction with a 1D detector array. Combining optical scanning and ghost imaging results in a novel 2D imaging technique that surpasses the capabilities of bulky crystals.

\begin{figure}[ht]
\centering
\includegraphics[width=0.95\linewidth]{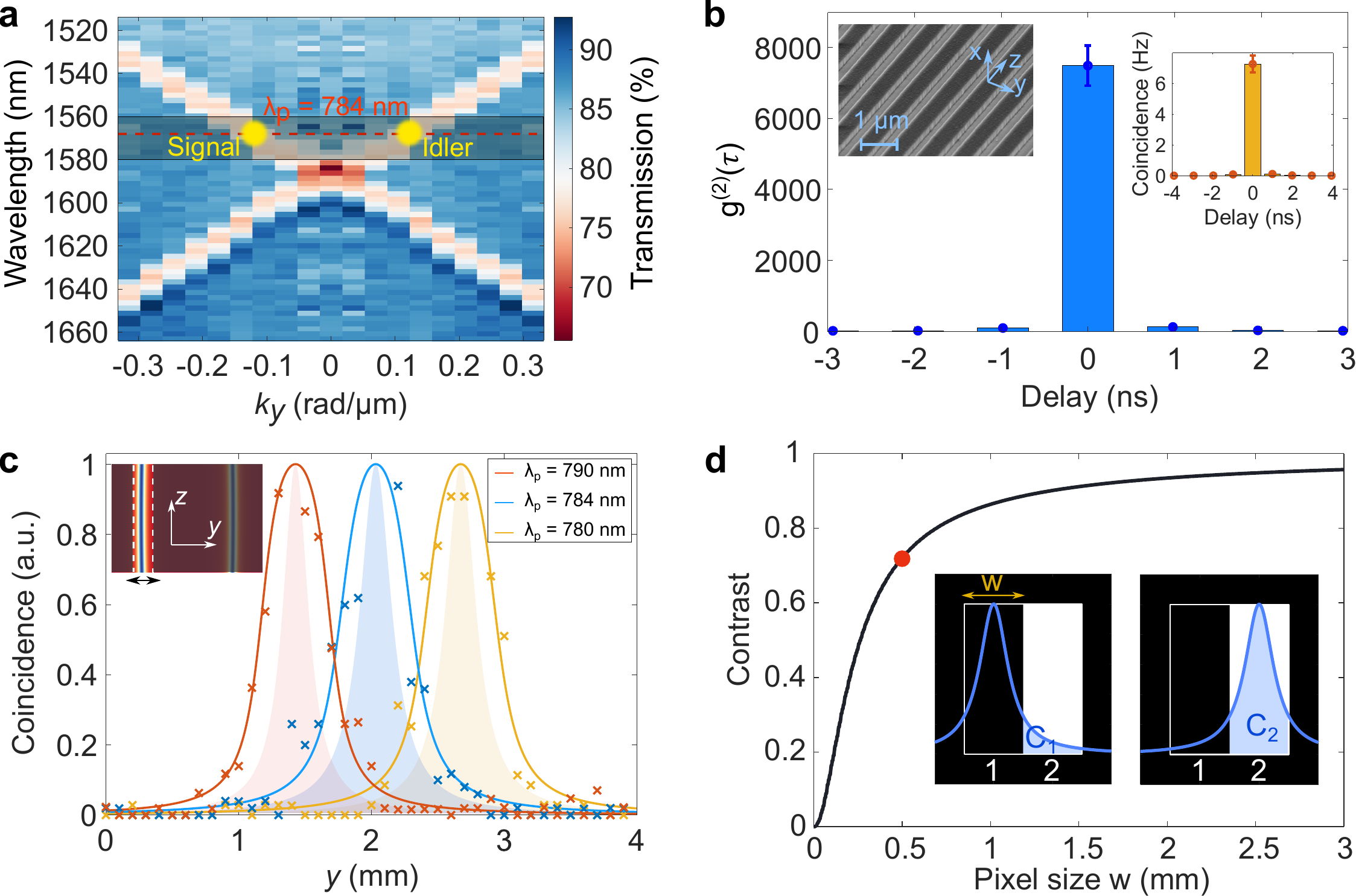}
    \caption{\textbf{Experimental calibration of photon-pair emission.} \textbf{a,} Linear transmission spectrum vs. wavevector component $k_y$. The metasurface manifests nearly degenerate optical resonances at 1584~nm for normal incidence and two-mode splitting at larger $k_y$. This feature enables linearly tunable photon emission in the grey-shaded region. %\comment
    {The transmission spectra are measured by tilting the metasurface along $k_y \ge 0$, while the spectra at $k_y<0$ are mirrored due to reflection symmetry to illustrate the transverse phase matching (i.e. $k_{y}^{(s)} \simeq - k_{y}^{(i)}$).} \textbf{b,} Second-order correlation function $g^{(2)} (\tau)$ of photon pairs from the metasurface. The photon correlation at zero delay exceeds the classical bound, i.e., $g^{(2)} (0) \gg 2$. The left inset shows the scanning electron microscope (SEM) image of the fabricated metasurface, incorporating an $x$-cut 300-nanometer-thick lithium niobate film with a $z$-oriented silica meta-grating sitting on top. The inset on the right presents the coincidence rate histogram of the photon pairs. %\comment
    {The error bars indicate one standard deviation.} \textbf{c,} Calibrated photon emission pattern along $y$-direction. The coincidence of photon pairs is measured by translating a $z$-oriented slit along $y$-direction (see inset). The crosses are the measured coincidence counts as a function of the slit position $y$ and the lines are the corresponding fitting under the assumption that the emission is a Lorenzian profile (shaded areas). Different colors represent different pump wavelengths which are marked with dashed lines in Fig.~\ref{fig:1}\textbf{c}. \textbf{d,}~Contrast of optical scanning imaging. For contrast analysis, we consider an object composed of two connecting pixels (each with a size $w$) of either transparent or opaque, as seen in the insets. The imaging contrast is evaluated via the total transmitted coincidence counts (blue-shaded areas $C_1$ and  $C_2$) at which the photon emission (blue curve) is scanned to the center of each pixel. The red dot ($w = 0.5$ mm) corresponds to the pixel size considered in the following imaging experiments.
    % \textbf{a}~Scanning Electron Microscope (SEM) image of the fabricated metasurface. The substrate of the metasurface is an $x-$ cut lithium niobate film with a 300~nm thickness. The silica linear metagrating added on top is oriented along the optical axis ($z$) of the film. \textbf{b}~ Experimental setup for optical scanning. The single photon detectors register the coincidence counts of the photon pairs from the metasurface, by translating a slit along $y$-direction in the signal arm. The same measurements are repeated at various pump wavelengths. \textbf{c}~Photon emission vs. pump wavelength. The emission pattern is calibrated from the coincidence measurements as a function of slit positions. The gap in wavelength tuning is attributed to the laser mode hopping (see Fig.4 in the Supplementary material). \textbf{d}~Imaging contrast of optical scanning imaging.  To analyze the resolution limit for distinguishing totally transparent and opaque pixels, we consider a 1D image composed of two connected pixels, each with a size $w$ (see insets). We evaluate the contrast of transmitted coincidence counts (blue-shaded areas) for which the emission pattern (blue curve) is tuned to the center of each pixel. The red dot corresponds to the pixel size considered in the following imaging experiments.
    }
    \label{fig:2}
\end{figure}

\subsection{Experimental characterization of photon-pair emission}
A scanning electron microscopy (SEM) image of the fabricated metasurface is shown in the inset of Fig.~\ref{fig:2}\textbf{b}, featuring a sub-wavelength-scale silica grating on top of an $x$-cut lithium niobate thin film of 300-nanometer thickness. {We tailor the angular dispersion to support one of the dual optical resonances at 1584~nm under normal incidence, by choosing a grating period (900~nm) and width (500~nm). These values are different from those of the geometry in Ref.~\cite{Zhang:2022-eabq4240:SCA}. The measured transmission spectra at different incident angles are shown in Fig.~\ref{fig:2}\textbf{a}.}
% Note that the geometry of the meta-grating is different from the one in~\cite{Zhang:2022-eabq4240:SCA}. By setting the period and width of the meta-grating as 900~nm and 500~nm respectively, we tailor specific angular dispersion with one of the dual optical resonances at 1584~nm under normal incidence, as shown in the measured transmission spectrum at different incident angles in Fig.~\ref{fig:2}\textbf{a}. 
This design allows us to linearly tune the emission angle of photon pairs generated from the metasurface, with a pump laser wavelength $\lambda_{\rm p}$ from 779~nm to 790~nm. As a result of the transverse phase matching condition $k_{y}^{(s)} \simeq - k_{y}^{(i)}$ in the same resonance band, the signal and idler photons are nearly degenerate at the wavelength $2\times\lambda_{\rm p}$, as indicated by the red dashed line. Correspondingly, the wavelength of photon pairs is tuned from 1558~nm to 1580~nm (marked in the grey-shaded area) by changing the pump wavelength.

At $2\times\lambda_{\rm p}= 1568$~nm as marked with the red dashed line in Fig.~\ref{fig:2}\textbf{a}, the second-order correlation function $g^{(2)} (\tau)$ of the photon pairs generated from the metasurface is shown in Fig.~\ref{fig:2}\textbf{b}. The value of $g^{(2)}$ at zero delays is $\approx$ 7000, substantially exceeding the classical bound of 2. The coincidence rate of $\approx$ 6 Hz shown in the right inset is measured with a pair of single-photon detectors based on InGaAs/InP avalanche photodiode at a detection efficiency of $25$\% for each detector, which can be substantially improved with superconducting nanowire single-photon detectors whose efficiencies typically exceed 90\%. Further improvement in the photon-pair rate can be achieved via material with higher second-order nonlinearity and improved design of the metasurface.

% resonance results in a 60-time enhancement in photon-pair generation rate compared to the unpatterned thin film. 

The emission pattern of photon pairs and its tunable feature are experimentally characterized in Fig.~\ref{fig:2}\textbf{c}, in which the photon coincidence is measured as a function of the position of a narrow aperture placed in the signal arm (see inset). The experimental procedures described here are different from the work that reported the tuning feature~\cite{weissflogDirectionally2024}, where the signal and idler photons were not separated and a beam block was used to scan through the emission. It is evident that the photon emission shifts along the $y$ direction as the pump wavelength changes. To avoid apparent cross-talk between pixels for the scanning imaging protocol, we select three pump wavelengths (780~nm, 784~nm, and 790~nm) at which the emission patterns of photon pairs have small overlaps. The selected wavelengths are unevenly distributed due to the mode hopping of the pump laser~(see supplement S3.4). We note that the number of pixels can be readily increased by using a laser with a more extensive wavelength tuning range, which can provide a substantial improvement in the field of view. 

% using the setup shown in Fig.~\ref{fig:2}\textbf{b}. A slit of 500 \textmu m width added in the signal arm is translated through the emission in the $y$-direction direction, while the idler photons are directly collected by the single-photon detector. The emission pattern at a given pump wavelength can be reconstructed by measuring the two-photon coincidence as a function of the silt position and fitting the results with a Lorenzian profile of photon emission (see supplement).  The discontinuity in pump wavelength is due to the mode hopping of the laser diode.

The resolution of scanning imaging is defined by the spatial bandwidth of photon emission (shaded area in Fig.~\ref{fig:2}\textbf{c}), which can be engineered through the quality factor of the optical resonances. For the spatial bandwidth calibrated in the experiment, we analyze the corresponding resolution by evaluating the imaging contrast of a 1D object composed of two connected pixels. As depicted in the insets of Fig.~\ref{fig:2}\textbf{d}, one of the pixels is totally transparent (white) while the other one is opaque (black), each with a width $w$. The emission pattern (blue curve) is scanned respectively to the center of each pixel by tuning the pump to two different wavelengths, at which the photon coincidence counts (blue areas) transmitted through the object are denoted as $C_1$ and $C_2$ correspondingly. The imaging contrast, defined as $(C_2-C_1)/(C_2+C_1)$, is shown in Fig.~\ref{fig:2}\textbf{d} as a function of the pixel width $w$. Our following experiments in optical scanning imaging will consider objects with a pixel size of 500 \textmu m (red dot), where the tuning range of pump wavelength can cover the scanning of an object composed of three connected pixels. Meanwhile, this pixel size guarantees a reasonable imaging contrast of above 70\%.

% As a demonstration of the proof of concept, we consider a binary object of three pixels with each pixel either fully transparent or opaque. To guarantee high-quality imaging, the rate of photon pairs transmitted through the object must have high contrast for different combinations of open or closed pixels. Considering a pixel size of 500 um, the expected SPDC rate normalized to the maximum is shown in Fig.~\ref{fig:2}\textbf{d}. The reconstruction error is defined as Contrast (open/close) as a function of pixel size. 

\subsection{Combined optical scanning and quantum ghost imaging}
%\subsection{Quantum imaging combining scanning and quantum ghost imaging}
\begin{figure}[ht]
\centering
\includegraphics[width=0.9\linewidth]{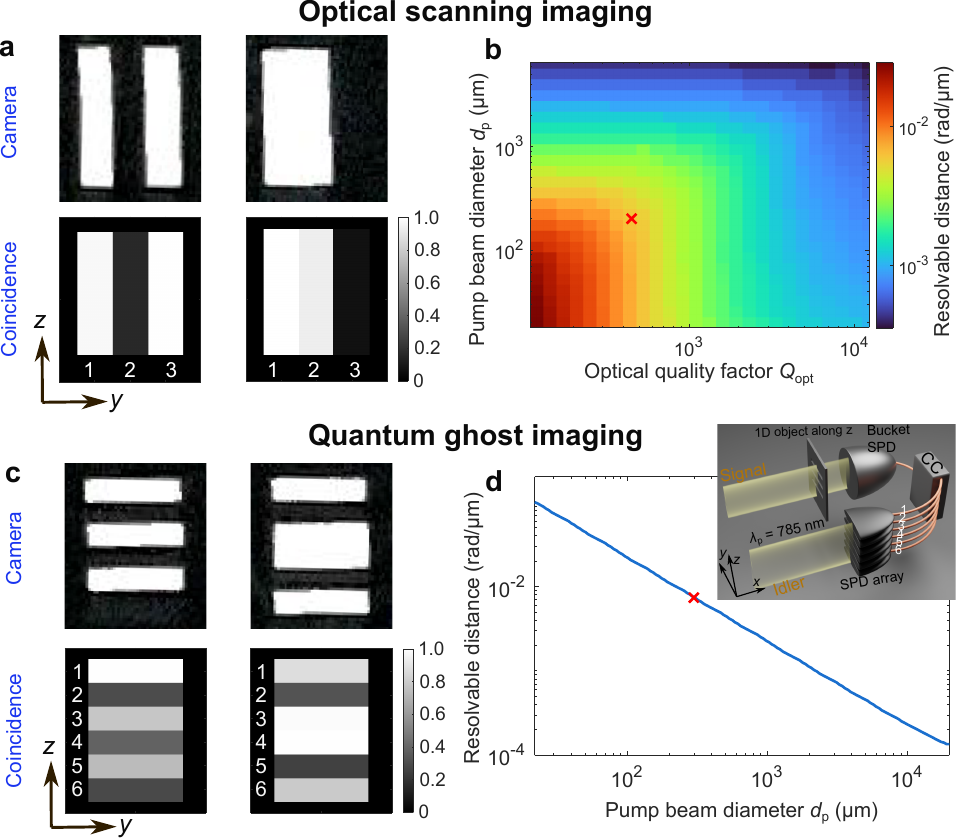}
    \caption{\textbf{Experimental characterization of separate imaging protocols.} \textbf{a,}~Experimental characterization of optical scanning imaging. A three-pixel 1D object along the $y$-direction is placed in the signal arm. The emission patterns of the photon pairs from the metasurface are scanned by tuning the pump wavelength to $\lambda_{\rm p} = $ 780~nm, 784~nm, and 790~nm, as considered in Fig.~\ref{fig:2}\textbf{c}. The normalized coincidence counts of photon pairs provide the object information. The top two figures are the optical camera images of the object and the bottom two are the corresponding coincidence measurements.
    \textbf{b,}~Simulated minimum resolvable distance of optical scanning imaging as a function of pump beam diameter $d_{\rm p}$ and quality factor $Q_{\rm opt}$ of the optical resonance. A higher $Q_{\rm opt}$ and larger $d_{\rm p}$ offer better imaging resolution. The red cross corresponds to the experimental conditions. 
    \textbf{c,} Experimental characterization of quantum ghost imaging. We consider two $z$-oriented objects composing six pixels of different patterns. The coincidence counts (normalized to the maximum) at each pixel correspond to the correlation measurements between the bucket and SPD array, containing information on the objects. The top and bottom figures display the camera images and coincidence measurements, respectively.
    \textbf{d,}~Simulated resolvable distance of quantum ghost imaging as a function of $d_{\rm p}$. The resolution of ghost imaging tends to approach the diffraction limit as the pump beam gets close to a plane wave while $Q_{\rm opt}$ is not relevant in this case. The red cross represents the experimental value of $d_{\rm p}$. The inset shows the setup for the 1D quantum ghost imaging of a 1D object placed in the signal arm. 
    }
    \label{fig:3}
\end{figure}

To understand the independent performance of optical scanning and quantum ghost imaging protocols, we initially perform the imaging just for 1D objects aligned with $y$ and $z$ directions. Both protocols are implemented in the Fourier plane of a 2-$f$ imaging system where the photon pairs are spatially anti-correlated. We note that the biphoton coincidences linearly correlate to the width and distance of the object pixels in the regime of interest that we realize experimentally (see supplement Figs.~S1\textbf{b,e} and~S2\textbf{b,e}). As a result, the coincidence measurements 
%presented below in this section 
directly provide the imaging data. For the quantum imaging results presented below in Figs.~3-4, the measurement time of coincidences is extended to 1~hour to improve imaging quality, under a pump power of 85 mW.

% The experimental setup for the optical scanning imaging is sketched in Fig.~\ref{fig:3}\textbf{a}, a bucket detector followed by the object is positioned in the signal arm while one element of the detector array collect the idler photons. 

For all-optical scanning imaging in the $y$-direction, we consider objects containing three binary pixels, with each pixel either fully transparent (white) or opaque (black), as shown in the top panels of Fig.~\ref{fig:3}\textbf{a}. The middle pixel is aligned with the peak position of the photon emission at a pump wavelength $\lambda_{\rm p} = 784$~nm. The coincidence counts of signal and idler photons are measured at three pump wavelengths specified in Fig.~\ref{fig:2}. The images of the objects are reconstructed from the coincidences normalized to the maximum, as provided in the bottom panels of Fig.~\ref{fig:3}\textbf{a}. The smallest contrast of black and white pixels is estimated as $\approx$71\%, which agrees with our estimation in Fig.~\ref{fig:2}\textbf{d}. We note that, unlike mechanical scanning~\cite{pittmanOptical1995}, this optical scanning protocol can be potentially ultra-fast with a frequency-comb pump~\cite{shaltoutSpatiotemporal2019}. 
%and ensure the stability of the imaging process.} 

Further, we analyze the potential for improving the imaging resolution in Fig.~\ref{fig:3}\textbf{b}. The resolution, defined as the minimum resolvable distance of two objects, is found to depend on the pump beam diameter $d_{\rm p}$ and quality factor $Q_{\rm opt}$ of the metasurface resonances. Since the object is positioned in the far field of photon emission, the imaging resolution is calculated in terms of transverse wavenumbers, in units of rad/\textmu m. The corresponding physical object dimensions can be flexibly selected by the focal length $f$ of the lens. For the lens used in experiments ($f = 50$ mm), 0.01~rad/\textmu m corresponds to 0.13~mm. We find that the spatial width of the photon-pair emission is inversely related to the spectral width of the optical resonance (as determined by $Q_{\rm opt}$) because of the conservation of energy and transverse momentum in the SPDC process. As a result, larger $Q_{\rm opt}$ leads to narrower spatial emission and better imaging resolution. Additionally, a larger pump beam diameter, which implies a smaller spatial bandwidth in $k$-space, results in a smaller $k$-space range for transverse phase matching conditions, a narrower region of photon emission, and an improved resolvable distance. The red cross in Fig.~\ref{fig:3}\textbf{b} indicates the experimental condition ($Q_{\rm opt} \approx 450$, $d_{\rm p} \approx 200$ \textmu m) described in Fig.~\ref{fig:3}\textbf{a}, in which the imaging resolution can be substantially improved using a laser with a larger pump beam diameter and a metasurface with a higher quality factor of the optical resonance. 

Figure~\ref{fig:3}\textbf{c} presents the quantum ghost imaging along the $z$ direction, using objects containing six pixels each of 350~\textmu m size (top panels). The inset of Figure~\ref{fig:3}\textbf{d} shows the schematic for this protocol. Signal photons transmitting through the object are collected by the bucket detector while idler photons are directly captured by the SPD array composed of six elements. At a fixed pump wavelength of 784~nm, we measure the correlation between the bucket detector and each element of the SPD array, based on which the images are obtained as shown in bottom panels in Fig. \ref{fig:3}\textbf{c}. We find that the imaging resolution is not dependent on the optical quality factor $Q_{\rm opt}$ due to the anti-correlation of photon pairs in the $z$-direction (see Fig.~\ref{fig:1}\textbf{e}). The pump beam diameter $d_{\rm p}$, however, determines the minimum resolvable distance, similar to conventional quantum ghost imaging~\cite{Padgett:2017-20160233:PTRSA}. The resolution can be improved for a larger $d_{\rm p}$ and approaches the diffraction limit as the pump beam is close to a plane wave. Consequently, we can obtain substantially better imaging resolution using a larger metasurface pumped with a plane-wave laser. In contrast, it is challenging to achieve ideal imaging resolution using photon pairs from bulky crystals {with a thickness larger than the coherence length}. {This is because bulky crystals are subject to longitudinal phase matching conditions~\cite{santiago-cruzEntangled2021e, OkothMicroscale2019, guoUltrathin2023} or technical limitations at larger apertures (such as with PPKTP crystals)~\cite{gilabertebassetPerspectives2019}}.

% The original image is reconstructed at a contrast of $\approx 40\%$.
% The contrast of the ghost imaging is \textcolor{red}{40\%}, which can be easily improved using a larger pump beam diameter and a larger metasurface.

% the schematic for the quantum ghost imaging along the $z$ direction. Signal photons transmitting through the object are collected by the bucket detector while idler photons are directly captured by the SPD array composed of six elements
% \subsection{2D quantum imaging and beyond} 

\begin{figure}[ht]
\centering
\includegraphics[width=1\linewidth]{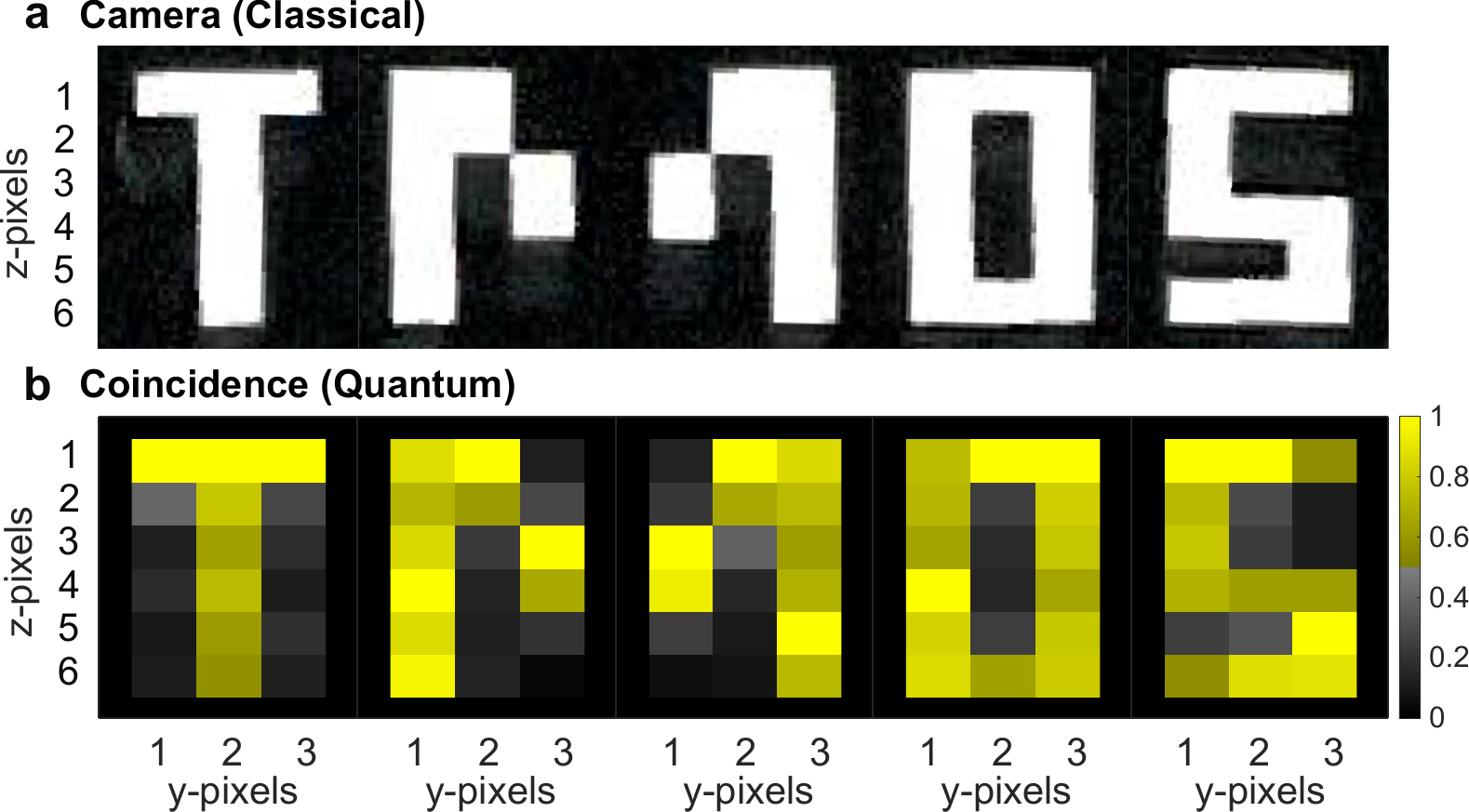}
    \caption{\textbf{2D quantum imaging combining ghost imaging and optical scanning.} \textbf{a,}~The optical camera images of objects. Each object contains patterns of different English letters, i.e., `T',`M',`O', and `S', which is composed of $3\times6$ pixels, with pixel dimensions of 500 mm $\times$ 350 mm. \textbf{b,}~Reconstructed images from coincidence measurements. We perform ghost imaging along $z$-direction at three wavelengths 780~nm, 784~nm, and 790~nm, providing comprehensive 2D image information. The coincidence count at each pixel is normalized to the global maximum at each pump wavelength. The image quality can be improved by processing the coincidence counts. Pixels with count values below 0.5 are rendered in a black-and-white color scheme, while pixels with counts above or equal to 0.5 are displayed in a yellow color scheme.
    }
    \label{fig:4}
\end{figure}

We now proceed with the analysis of quantum imaging of 2D objects, combining the scanning imaging along $y$ and ghost imaging along $z$-direction. We consider objects of 3$\times$6 pixels, with each pixel size of 500 \textmu m $\times$ 350 \textmu m, as shown in Fig.~\ref{fig:4}\textbf{a} for the camera image of the objects. We perform quantum ghost imaging at three different pump wavelengths, with each one corresponding to the ghost imaging information of one column of the object along $z$. The normalized coincidence measurements for each pixel are shown in Fig.~\ref{fig:4}\textbf{b}, where the images of the objects are clearly seen.  As we have the a priori knowledge of the object pixels being either totally transparent or blocked, we can further process the image by checking through the coincidence of each pixel and setting a yellow colour scheme if the normalized coincidence is larger than~0.5. We find that the 2D image can be fully reconstructed with a 100\% success rate. 

We perform a proof-of-principle experiment, where each object incorporates only $3\times6$ pixels due to the limitations of laser tunability and numerical aperture of fibers (for photon collection). However, the number of resolution cells can be readily extended even beyond what has been achieved with bulky crystals. The cell number is quantified by the ratio of the field of view to the minimum resolvable distance. For quantum ghost imaging with bulky crystals, the number of cells $\mathcal{N}_{\rm bulky}$ is limited by the pump beam diameter $d_{\rm p}$ and the crystal length $L$, i.e., $\mathcal{N}_{\rm bulky}=\pi d_{\rm p}^2/16L \lambda_{\rm p}$~\cite{moreauResolution2018a}. Typical values of $d_{\rm p}=1$ mm and $L=$ 1 mm used in previous works~\cite{sephtonRevealing2023,moreauResolution2018a, aspdenPhotonsparse2015} lead to $\mathcal{N}_{\rm bulky}\approx10^3$ for $\lambda_{\rm p} = 780$~nm. In contrast, our protocol in the $y-$direction has no potential restriction for the field of view and the predicted minimum resolvable distance (see Fig.~\ref{fig:3}\textbf{b}) is $\approx 0.1$ mrad/\textmu m, given an experimentally achievable quality factor $Q_{\rm opt}\approx1000$~\cite{SongNonlocal2021} and pump beam diameter $d_{\rm p}=2$ cm (as allowed by a centimeter-scale metasurface). This translates to the pixel number of $\mathcal{N}_{\rm meta,y}\approx4\times10^4$ assuming a photon collection angle of $\pm30$ deg (i.e., $k_y\in[-2, 2]$ rad/\textmu m). For the $z-$direction, the $k_z$ range for a continuous emission is $\approx 0.2$ rad/\textmu m and the minimum resolvable distance for $d_{\rm p}=2$ cm is $\approx 0.1$ mrad/\textmu m (see Fig.~\ref{fig:3}\textbf{d}), corresponding to the pixel number of $\mathcal{N}_{\rm meta,z}\approx2\times10^3$. As a result, our protocol can potentially achieve a total number of resolution cells $\mathcal{N}_{\rm meta}= \mathcal{N}_{\rm meta,y}\mathcal{N}_{\rm meta,z}\approx 8\times10^7$, which can be about five orders of magnitude larger than what was possible with bulky crystals. 
%---------------------------------------------------
\subsection{Multi-wavelength imaging and beyond}
\begin{figure}[ht]
\centering
\includegraphics[width=0.95\linewidth]{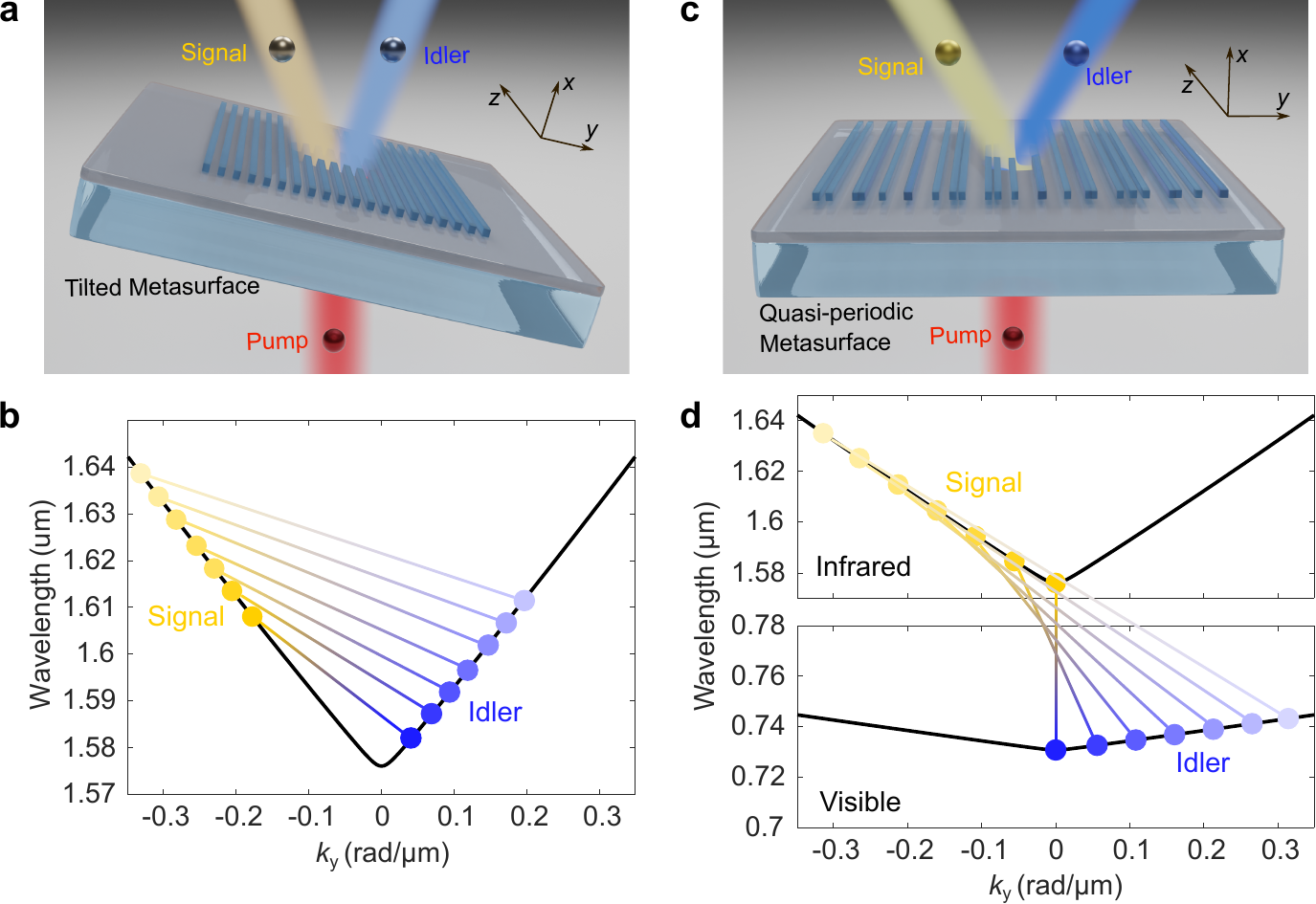}
    \caption{\textbf{Quantum imaging using non-degenerate photon pairs.} \textbf{a,} Generation of non-degenerate photon pairs from a tilted metasurface. As the incident pump beam is off-normal to the metasurface along the $y-$direction, the metasurface produces photon pairs non-degenerate in wavelength. \textbf{b,} Same-band phase matching for the non-degenerate SPDC. Tilting the metasurface introduces a non-zero transverse wavevector for the pump beam, resulting in asymmetric photon emission across the grating and non-degenerate SPDC. Tuning the pump wavelength shifts the emission angles of both signal (yellow dot) and idler (blue dot) photons, as manifested in different colour gradients. \textbf{c,} Non-degenerate SPDC with a quasi-periodic metasurface. A metagrating incorporating two periods can create double optical resonances at both visible and infrared wavelengths, allowing the enhanced generation of photon pairs highly non-degenerate at corresponding resonance wavelengths. \textbf{d,} Cross-band phase matching for photon pairs at optical and infrared domain. Conservation of energy and transverse momenta can be achieved across the visible and infrared resonance bands, allowing the generation of infrared signal photons and visible idler photons. The emission angles are also optically tunable via pump wavelength, which can be used for combined quantum imaging with signal photons illuminating the object and idler photons collected with a single-photon camera available in the optical domain.}  
    \label{fig:5}
\end{figure}
    %The signal photons emitted from the metasurface are bounced back from a moving object and are then reflected by the same metasurface. The reflected signal photons combined with idler photons are collected with a fiber collimator and are then split into two optical paths using a fiber beam splitter (FBS). The signal and idler photons are collected by a pair of SPD. The real coincidence of the photon pairs is analyzed using the coincidence counter, which is further processed with a PID controller and fed to the tunable laser for controlling the laser wavelength and thereby the photon emission. Tracking the object is realized by finding the maximum of coincidence counts.

There has been extensive interest in using non-degenerate photon pairs for quantum ghost imaging~\cite{gilabertebassetPerspectives2019, aspdenPhotonsparse2015}, where the signal photons at the wavelength of interest illuminate the object and the idler photons at the visible wavelength are collected using single-photon cameras available in the optical domain. This approach enables the imaging at the wavelength range where the single-photon camera or detector arrays are impractical or inefficient. Although the quantum imaging protocol combining optical scanning and ghost imaging is focused on the degenerate regime in this work, the concept can be extended to the non-degenerate cases using the approaches sketched in Fig.~\ref{fig:5}. By simply tilting the metasurface (Figs.~\ref{fig:5}\textbf{a-b}), photon pairs with wavelength differences can be generated because the non-zero transverse wavevector $k_{\rm p, t}$ of the pump beam contributes to the transverse phase matching condition, i.e., $k_{\rm s, t}+k_{\rm i, t}= k_{\rm p, t}\neq 0$ (where $k_{\rm s, t}$ and $k_{\rm i, t}$ are the transverse wavevectors of the signal and idler photons, respectively). Importantly, the optical scanning protocol is still accessible by tuning the pump wavelength. The non-degeneracy achieved in this approach is subject to the tilting angle and is potentially limited by the angular dispersion of the optical resonance supported by the metasurface. In an alternative approach, highly non-degenerate photon pairs can be produced using a quasi-periodic metasurface supporting dual optical resonances at both visible and infrared (or at other on-demand regions of the spectrum) wavelengths, as shown in Fig.~\ref{fig:5}\textbf{c}. SPDC process can happen efficiently following the conservation of energy and momentum across two resonance bands (Fig.~\ref{fig:5}\textbf{d}). Again, the emission angles of photon pairs can be tuned optically via the pump wavelength. As a result, the non-degenerate photon pairs with optically tunable emission from the metasurfaces can be leveraged to realize the imaging protocol developed in this work, offering new opportunities for multi-wavelength imaging. Compared to the classical multi-colour ghost imaging requiring a spatial light modulator or other bulky components~\cite{chanTwocolor2009,duanNondegenerate2019}, this metasurface-based protocol ensures ultra-compactness with no need for additional components.

Additionally, the optical beam steering offers outstanding potential for LiDAR (light detection and ranging), optical communications, and imaging applications. State-of-the-art technologies for beam steering primarily rely on mechanical means or liquid crystals~\cite{beriniOptical2022}. Here we have shown that the photon emission direction can be optically tuned via the pump laser wavelength in a continuous $k$-space. While the nonlinear metasurface is passive, this optical tuning can be potentially ultra-fast with a frequency-comb pump~\cite{shaltoutSpatiotemporal2019}. 
Such a quantum beam steering unattainable with conventional bulky nonlinear crystals offers novel opportunities for quantum LiDAR~\cite{lopaevaExperimental2013a, liuEnhancing2019, Mrozowski:2024-2916:OE, zhaoLight2022}, quantum communications~\cite{lopaevaExperimental2013a}, and other contexts. For example, quantum characters of light, such as entanglement or strong timing correlation of photon pairs, have been used to improve the signal-to-noise ratio (SNR) in LiDAR via quantum illumination~\cite{lloydEnhanced2008, lopaevaExperimental2013a} and correlation measurement. Furthermore, our results suggest a concept of ``quantum object tracking'', in which a moving object can be actively tracked using entangled photon pairs from the metasurface via feedback control of the pump wavelength (see the sketch in Fig.~S7 in the supplement). We expect that this unique feature of quantum light with beam steering enabled by the nonlocal metasurface will open the door for novel sensing and imaging applications.

\section{Conclusions}
In this work, we reveal the first practical application that can benefit from the unique advantages of metasurface-based photon pair sources. We propose theoretically and demonstrate through proof-of-principle experiments quantum imaging combining optical scanning and ghost imaging, using spatially entangled photon pairs from a lithium niobate nonlinear metasurface where a linear silica grating is on top of a subwavelength-thick lithium niobate film. 
% The metasurface supports nonlocal resonances featuring distinct angular dispersion between parallel ($z$) and orthogonal ($y$) to the grating orientation, allowing continuous emission along $z$-direction and narrow emission along $y$-direction with emission angle optically tunable via pump wavelength.
% We perform 1D imaging for either direction separately: (i) optical scanning imaging along $y$ by tuning the pump wavelength; and (ii) quantum ghost imaging along $z$. We then demonstrate 2D imaging of several object patterns by combining these two protocols. 
The 2D images reconstructed via coincidence measurements show good agreement with those taken with an optical camera. The processed image reconstructs the original image with 100\% sucessbililty.  

We anticipate that our protocol will establish a foundation for the development of novel quantum imaging applications employing metasurfaces-based quantum light sources, with potential advantages over conventional quantum ghost imaging including (i)~A larger field of view. Along the $y$-direction, the photon emission angle can be tuned to cover almost the full $k$-space. Along $z$, the longitudinal phase matching condition is effectively removed such that photon emission is continuous over a very broad angle range. (ii)~Higher imaging resolution. The present fabrication techniques such as optical lithography allow the fabrication of metasurface up to a few centimetres, which would enable a larger pump beam diameter and dramatically improved imaging resolution. (iii)~Ultra-compact integrated devices. The thickness of the nonlinear metasurface is subwavelength at the nanometer scale, which is three to four orders of magnitude thinner than conventional bulky crystals. (iv)~Optical scanning imaging and beyond. The metasurfaces with nonlocal optical resonances feature all-optically tunable emission of spatially entangled photons in both degenerate and non-degenerate wavelengths. Such beam steering may allow novel sensing and imaging applications such as quantum LiDAR or object tracking. The unique flexibility enabled by metasurface such as polarization, spectral and spatial engineering can be further introduced to enrich the imaging data. (v)~Implications for other imaging techniques. Since the ultra-thin metasurface enables high degrees of spatial entanglement, this combined imaging protocol can potentially be extended to realize phase imaging~\cite{sephtonRevealing2023}, fractional Fourier imaging~\cite{weimannImplementation2016} and more.

The next essential step towards practical quantum imaging application with metasurface is to enhance the generation rate of photon pairs, which can be significantly improved by using materials with higher second-order nonlinear coefficients such as III\nobreakdash-V semiconductor materials~\cite{sultanovFlatoptics2022}, and 2D materials~\cite{guoUltrathin2023}, and by designing triple resonances at the signal, idler and pump photons, combined with the angular dispersion engineering. Once the photon-pair rate becomes comparable to bulky crystals, quantum imaging using photon pairs from metasurfaces will provide brand new possibilities and pave the way for the advancement of quantum imaging technologies.

%%% References

\section{Methods}
% \subsection{Numerical simulations}
% The photon emission patterns, spatial correlation of photon pairs, and imaging resolution are simulated and analyzed with coupled-mode theory in MATLAB. The experimental characterization of emission patterns and imaging reconstruction based on correlations are also performed in MATLAB.

\subsection{Fabrication of metasurface}
The metasurface fabrication started from a lithium niobate film of 303.7 nm thickness on a quartz substrate (NANOLN). After ultrasonic cleaning, the sample was deposited with a thin SiO$_2$ layer ($\sim$180~nm thick) using plasma-enhanced chemical vapor deposition. Next, PMMA was spin-coated to serve as the resist for electron beam lithography, after which a 30~nm thick nickel layer was deposited via electron beam deposition. The subsequent lift-off process leads to a nickel pattern on top of the SiO$_2$ layer, acting as a mask for etching the SiO$_2$ layer through inductively coupled plasma etching. Finally, the remaining nickel mask was removed with chemical etching. The fabricated metasurface has the size of 400 \textmu m $\times$ 400 \textmu m. 

% \comment{The designed metasurface was fabricated in a cleanroom starting from a LiNbO$_3$ film on a quartz substrate from NANOLN. After ultrasonic cleaning, a thin layer of SiO$_2$ with a thickness of $\sim$200~nm was deposited on the sample by plasma-enhanced chemical vapor deposition. Then, PMMA was spin-coated as the resist for the subsequent electron beam lithography. After lithography, a thin layer of nickel with a thickness of 30~nm was coated by electron beam deposition followed by a lift-off process. The nickel pattern was used as the mask for the etching of the SiO$_2$ layer by inductively coupled plasma etching. Finally, the residual nickel mask was removed by chemical etching. The size of the grating is 400 \textmu m $\times$ 400 \textmu m.}

\subsection{Experimental setup}
The laser (FPL785P, Thorlabs) for pumping the metasurface features a tunable wavelength from 779~nm to 791~nm. The fluorescence %carried by 
co-propagating with the incoming pump beam is suppressed by a short-pass filter at 850~nm. A lens with a focal length of 200~mm is used to focus the pump beam to a beam diameter of 200~\textmu m on the metasurface, followed by a lens with a focal length of 50 mm for collimating the photon pairs emitted from the metasurface. To perform high-quality quantum measurements, the pump laser beam and fluorescence from the metasurface and other optics are filtered with a long-pass filter at 1100~nm and a band-pass filter at 1570~nm with a 50~nm bandwidth. The collimated photon pairs with a high coincidental-to-accidental ratio are spatially split into two optical paths with a $D$-shaped mirror. One optical path includes a scanning slit mimicking the 1D detector array while the other path contains objects for imaging. The photons in the two paths are respectively collected with multimode fibers, which are then sent to single-photon detectors based on InGaAs/InP avalanche photodiodes (ID230, IDQ). The coincidence events in timing correlations are characterized by a time-to-digital converter (ID801, IDQ), in which the coincidence window is set at 0.486~ns.

\subsection{Fabrication of object} 
The fabrication of the designed objects for imaging started with the patterning of photoresist (AZ1512H, MicroChemicals) using mask writer (SF100 XPRESS, Intelligent Micropatterning) on a 4-inch glass wafer. Then a thin chromium layer with a thickness of 200 nm was deposited on the wafer by electron beam evaporation (Nanochrome II, Intlvac) followed by a lift-off process. The wafer was diced into samples of 15 mm  $\times$ 15 mm using a wet dicing saw (DISCO 3350).

\bmhead{Data availability}
All data needed to evaluate the conclusions in the paper are present in the paper and/or the Supplementary Materials.
\small

\bmhead{Acknowledgments}
This work was supported by the Australian Research Council (DP190101559, CE200100010). The nonlinear metasurface fabrication was performed at the Australian National University node of the Australian National Fabrication Facility (ANFF), a company established under the National Collaborative Research Infrastructure Strategy to provide nano and micro-fabrication facilities for Australia’s researchers. This work was also performed in part at the Melbourne Centre for Nanofabrication (MCN) in the Victorian Node of the Australian National Fabrication Facility (ANFF).
\small

\bmhead{Author contributions}
A.A.S. and J.Ma conceived the idea, designed the research, and supervised the project. J.Ma developed the experimental setup. J.Ma, J.R. and C.M.B. performed the experiments and processed the data. J.R., J.Ma, J.Z., and A.A.S. performed theoretical analysis. J.Z, J.Meng, and K.B.C. fabricated the metasurface and objects. J.Ma wrote the paper. All authors discussed the results and commented on the paper.
\small

\bmhead{Competing interests}
The authors declare no competing interests.
\small

\bibliography{MSimaging,mybib2}

\end{document}